# Peculiarities of timing and spectral diagrams of magnetic video-pulse excitation influence on NMR spin-echo in magnets

G.I. Mamniashvili[a], T.O. Gegechkori[a], A.M. Akhalkatsi[b], T.A. Gavasheli[b],

E.R. Kutelia[c], L.G. Rukhadze[c], D.I. Gventsadze[c]

[a]*Andronikashvili Institute of Physics of I.Javakhishvili Tbilisi State University,*
 *6 Tamarashvili str., 0177. Tbilisi, Georgia.*
[b]*I.Javakhishvili Tbilisi State University, 3 Chavchavadze av. 0128. Tbilisi, Georgia.*
[c]*Georgian Technical University, 69, Kostava str. Tbilisi, 0175, Georgia*

zviadadzemichael@yahoo.com

## Abstract

We present the first systematic study of timing and spectral diagrams of magnetic video-pulse influence on the NMR two-pulse echo in a number of magnets (ferromagnets, ferrites, half metals, intermetals). It is shown that the timing diagrams showing the dependence of two-pulse echo intensity on the temporal location of a magnetic video-pulse in respect to radio-frequency pulses and the spectral diagrams of this influence are defined mainly by the local hyperfine field anisotropy and domain walls mobility. These diagrams could be used for the identification of the nature of NMR spectra in multidomain magnetic materials and to improve the resolution capacity of the NMR method in magnets.

## 1. Introduction

The possibilities of using different methods of nuclear spin-echo spectrometry for studying properties of magnetically ordered substances were analyzed in a large number of works [1]. One of such widely employed methods is based on the introduction of additional pulses of a dc magnetic field into the system of exciting radio-frequency (RF) pulses; these pulses were called magnetic video-pulses (MVPs), since they lack the filling frequency. Thus, in [2-4] the MVPs were used to investigate the properties of domain walls (DW) in the europium iron garnet $Eu_3Fe_5O_{12}$, in ferrites with a spinel structure, thin magnetic films and $Y_{2-x}Gd_xCo_{17}$ compounds with the substitution of Gd for Y ions. In these works, the different role of MVPs influence during their symmetrical and asymmetrical position with respect to the second RF pulse in the two-pulse echo (TPE) procedure was revealed. These differences make it possible to find the coercive-force-related distribution of DW upon the symmetrical arrangement of MVPs, and the anisotropy of the hyperfine field (HF) at the nuclei in the case of its asymmetrical arrangement. In particular, the detection [3] of the inhomogeneous influence of MVPs on different sections of the frequency spectrum of $^{59}$Co NMR in $(Y_{0.9}Gd_{0.1})_2Co_{17}$ made it possible to reveal those crystallographic positions that prove to be preferable upon the substitution of Gd for Y.

Previously, such a procedure was used [2] for determining the magnetic-field strength which shifts a DW by a distance equal to its thickness. The scheme of the experiment is shown in Fig. 1, borrowed from [4].

The authors of [4] investigated the influence of MVPs on the intensity of TPE. They showed that due to comparatively small hyperfine field anisotropy on $^{57}$Fe in lithium ferrite the maximum suppression effect is achieved when the MVP coincides with one of the RF pulses. This is caused by the fact that the action of a MVP on the multidomain ferromagnetic material is in essence reduced to a reversible (in weak magnetic fields) displacement of DW. A MVP symmetric in respect to the second RF pulse has the effect that both the first and second RF pulses excite nuclei whose positions within the DW change. These positions define the resonance frequency as well as the factor of enhancement of the RF field $\eta$ inside DW [1]. If the HF field inhomogeneity is small and the excitation pulses are sufficiently short then the change of resonance frequency can be neglected. In this case the change in the RF enhancement factor $\eta$ reduces the echo intensity [2], as it is seen for lithium ferrite, Fig. 1. The application of a MVP asymmetrically in the interval between the RF pulses can significantly influence the intensity of TPE only in the case of anisotropic HF interactions [2].

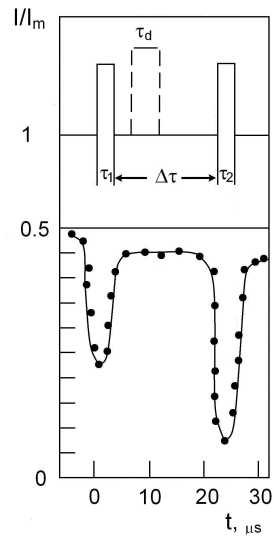

Fig. 1. Timing diagrams of the relative intensity $I/I_{max}$ dependence of the two-pulse echo (•) on the temporal location of a MVP of $H_d$=5 Oe, for $^{57}$Fe NMR in lithium ferrite at: $\tau_1 = \tau_2 = 0.8$ μs, $\Delta\tau = 21$ μs, $\tau_d = 3$ μs, $f_{NMR}$=74.0 MHz, $H_d$=0. $\tau_1$, $\tau_2$, $\Delta\tau$, $\tau_d$ are rf pulse durations, time interval between them and magnetic pulse duration, correspondingly.

So, the great importance is such characteristics of magnetic materials as the mobility of DW and HF based anisotropy, which are different in magnetically soft cubic lithium ferrite and magnetically harder material as uniaxial cobalt.

## 2. Experimental results and their discussion

To study the above noted properties of the magnetic materials in more detail, we carried out experimental concerning the effect of a MVP with an amplitude of the magnetic field up to $H_d$=500 Oe and durations equal to several microseconds on the signals of TPE in several magnetic substances, namely, polycrystalline cobalt-thin magnetic films (TMF), half metal $Co_2MnSi$, Co-Cu and MnSb ferromagnetic alloys. The half metals are regarded as promising materials for spintronics [5]. The NMR spectrometer and MVP excitation technique, as well as the procedures of sample preparations are described in [4, 6-9].

As it follows from Figs. 1-7, the essential difference is seen for MVP action on the TPE in different magnets. The dependences of TPE intensities at application of a MVP for $^{59}$Co echo signal in Co TMF (Fig. 2), $^{59}$Co echo in Co-Cu (Fig. 3) and $Co_2MnSi$ (Fig. 4,5a, 6a) differs considerably from closer-to-each-other dependences for $^{57}$Fe echo in lithium ferrite (Fig. 1) and $^{55}$Mn echo in $Co_2MnSi$ (Fig.5b, 5c, 6b) and MnSb (Fig.7).

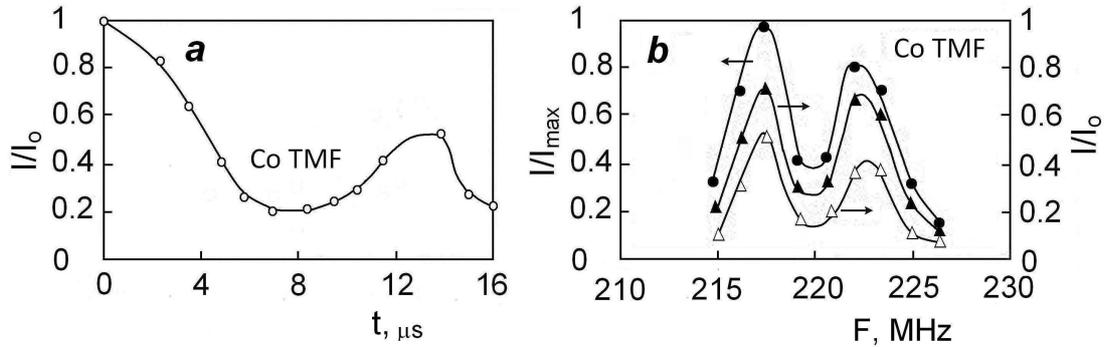

Fig. 2. Timing diagrams of the intensity dependence of the two-pulse echo on the temporal location of a MVP with duration $\tau_d$ and amplitude $H_d$ in

(*a*) cobalt thin magnetic films at: $\tau_1 = \tau_2 = 1.5$ μs, $\Delta\tau = 9$ μs, $\tau_d = 3$ μs, $H_d$=10 Oe, $f_{NMR}$=216.5 MHz, $I_o$ – echo amplitude at $H_d$=0.

(*b*) NMR spectrum of Co film (•) and frequency spectra diagrams for MVP influence for symmetric (▲) and asymmetric (△) application at: $\tau_1 = 1.3$ μs, $\tau_2 = 1.5$ μs, $\Delta\tau = 9$ μs, $\tau_d = 3$ μs, $H_d$=10 Oe.

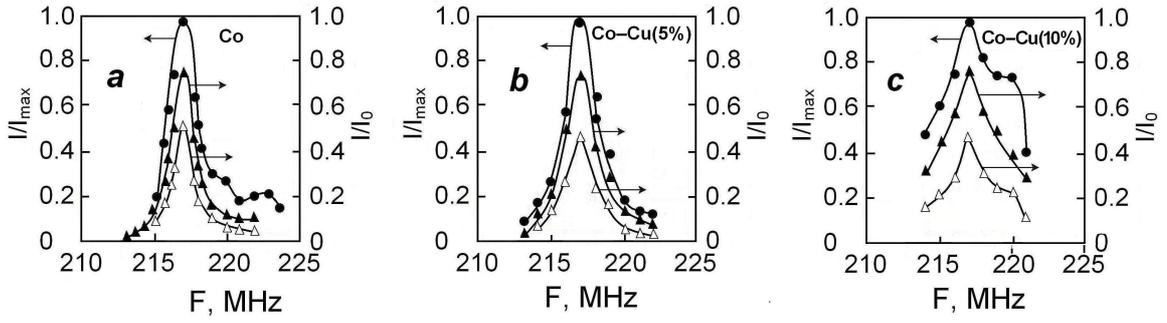

Fig. 3. NMR spectra (•) for Co (*a*) and Co-Cu alloys (*b,c*) for comparisons (b – 5 % Cu, c – 10 %) and spectral diagrams for MVP influence in case of symmetric (▲) and asymmetric (△) MVP application with amplitude $H_d$=350 Oe.

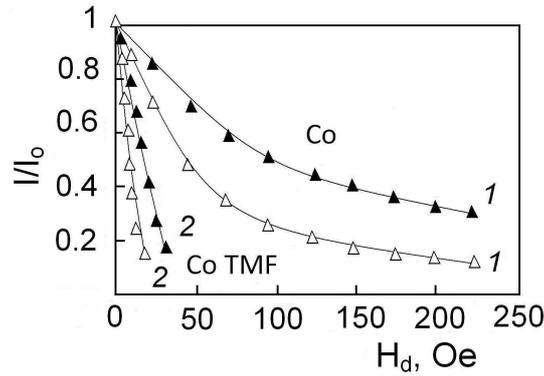

Fig. 4 Two-pulse echo intensity amplitude dependence on magnetic video-pulse amplitude (▲) and asymmetric (△) influence in polycrystalline cobalt at: $\tau_1$ = 1.3 μs, $\tau_2$ = 1.5 μs, $\Delta\tau$ = 9 μs, $\tau_d$ = 3 μs, $f_{NMR}$=218 MHz (*1*) and cobalt thin magnetic films at: $\tau_1$ = 1.3 μs, $\tau_2$ = 1.5 μs, $\Delta\tau$ = 9 μs, $\tau_d$= 3 μs, $f_{NMR}$=218 MHz (2).

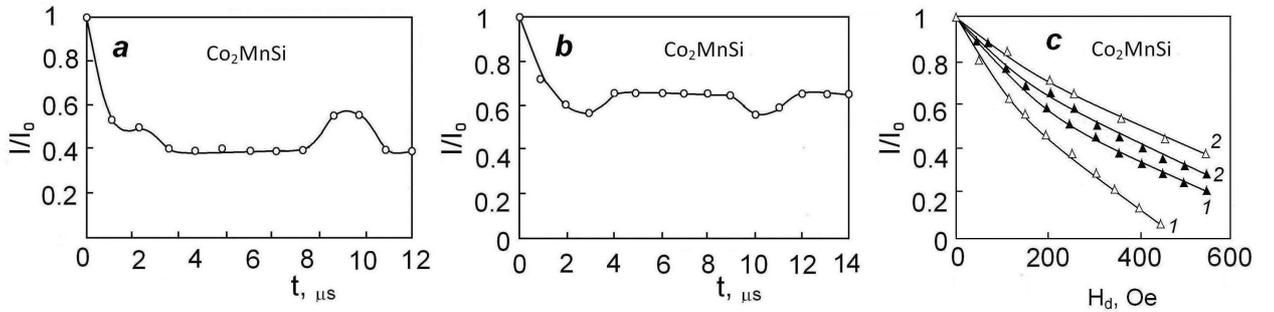

Fig. 5. Timing diagrams of the intensity dependence of the two-pulse echo on the temporal location of a MVP with duration $\tau_d$ and amplitude $H_d$ in $Co_2MnSi$

(*a*) for $^{59}$Co NMR at: $\tau_1$=1.1 μs, $\tau_2$=1.4 μs, $\Delta\tau$=10 μs, $\tau_d$=2 μs, $f_{NMR}$=145.5 MHz, $H_d$=550 Oe;

(*b*) for $^{55}$Mn NMR at: $\tau_1$=$\tau_2$=3 μs, $\Delta\tau$ = 7 μs, $\tau_d$ = 2 μs, $H_d$=300 Oe, $f_{NMR}$=354 MHz, $I_o$ – echo amplitude at $H_d$=0.

(*c*) amplitude diagrams of the TPE intensity dependence on the MVP value with duration $\tau_d$ and amplitude $H_d$ in $Co_2MnSi$ for symmetric (▲) and asymmetric (△) influence for:

1) $^{59}$Co NMR spin echo at: $\tau_1$ = $\tau_2$ = 2 μs, $\Delta\tau$ = 10 μs, $\tau_d$ = 3 μs, $f_{NMR}$=145 MHz.

2) $^{55}$Mn NMR spin echo at: $\tau_1$ = 0.8 μs, $\tau_2$ = 0.9 μs, $\Delta\tau$ = 8 μs, $\tau_d$ = 1.6 μs, $f_{NMR}$=353 MHz, $I_o$ – echo amplitude at $H_d$=0.

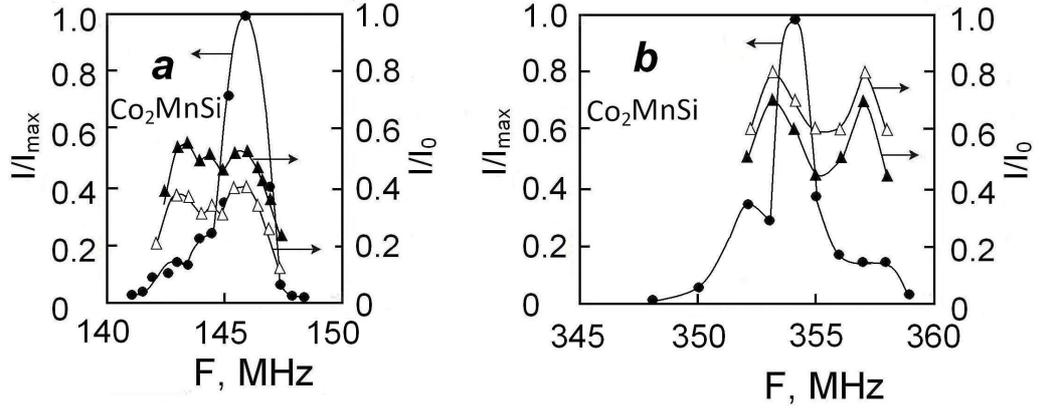

Fig. 6. Frequency dependence of the effect of the symmetric (▲) and asymmetric (△) MVP on the two-pulse echoes intensities in $Co_2MnSi$ for $^{59}Co$ NMR (*a*) and $^{55}Mn$ NMR (*b*) at:

(*a*) $\tau_1 = 1.1$ μs, $\tau_2 = 1.4$ μs, $\Delta\tau = 10$ μs, $\tau_d = 2$ μs, $H_d=550$ Oe;

(*b*) $\tau_1 = 0.8$ μs, $\tau_2 = 1$ μs, $\Delta\tau = 13$ μs, $\tau_d = 4$ μs, $H_d=190$ Oe; $I_o$ – echo amplitude at $H_d=0$.

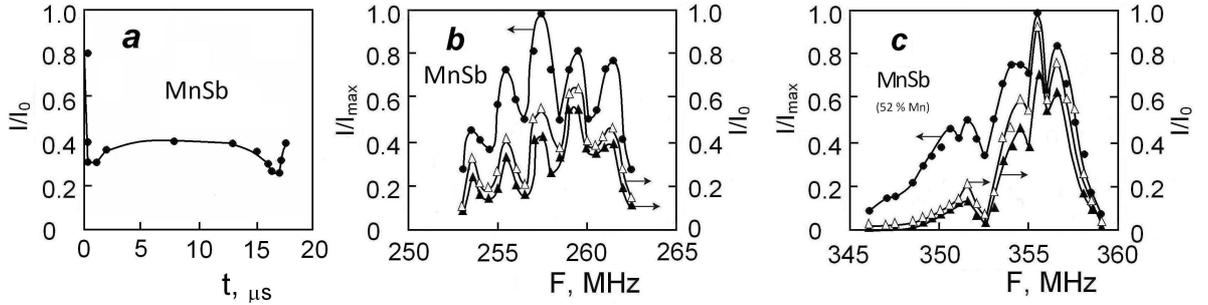

Fig. 7 (*a*) Timing diagrams of the intensity dependence of the two-pulse echo on the temporal location of a MVP of $H_d$, lasting for a time interval $\tau_d$ in MnSb at: $\tau_1 = 1.2$ μs, $\tau_2 = 1.6$ μs, $\Delta\tau = 17$ μs, $\tau_d = 2$ μs, $H_d=280$ Oe, $f_{NMR}=257$ MHz. $I_o$ – echo amplitude at $H_d=0$.

NMR spectrum of MnSb (•) and spectral diagrams of MVP influence for symmetric (▲) and asymmetric (△)) application for (*b*) $Mn_{0.5}Sb_{0.5}$ at: $\tau_1 = 1.2$ μs, $\tau_2 = 1.5$ μs, $\Delta\tau = 10$ μs, $\tau_d = 1.6$ μs, $H_d=280$ Oe and (*c*) for $Mn_{0.52}Sb_{0.48}$ at: $\tau_1 = 0.4$ μs, $\tau_2 = 0.5$ μs, $\Delta\tau = 10$ μs, $\tau_d = 1.6$ μs, $H_d=90$ Oe.

The influence of MVPs on echo intensity along the NMR frequency spectra was firstly studied in [2,3].

In Fig. 2 we present the timing and spectral diagrams of MVP influence on the $^{59}Co$ TPE in TMF cobalt samples. The MVP spectral influence on $^{59}Co$ in $Co_{1-x}Cu_x$ system (x=0.5 %, 10 %) is given in Fig. 3. Corresponding MVP influence dependences for TPE in $Co_2MnSi$ and MnSb samples are presented in Figs 5-7. Besides it, in Fig. 4 and Fig. 5c it is presented the amplitude dependences of TPE intensities for $^{59}Co$ and $^{55}Mn$ spin echoes on the amplitude of MVP $H_d$ in Co and TMF, and $Co_2MnSi$, correspondingly.

Note the difference in shape of the timing and order of spectral diagrams of MVP influence for the $^{59}$Co and $^{55}$Mn nuclei in Co$_2$MnSi (Fig. 5,6), correspondingly. This apparently reflects difference in the anisotropies of HF fields for these nuclei what affects the dependences of the intensity of the TPE on the time of application of MVPs: the symmetric MVPs which coincide in time with the RF pulses comparatively less suppress the signals of echo by $^{59}$Co nuclei and comparatively more reduce TPE of $^{55}$Mn nuclei as compared with asymmetric MVPs. So, the type of timing or spectral diagrams is defined mainly by the HF field anisotropy of corresponding nuclei which is small for $^{57}$Fe and $^{55}$Mn as compared for that of $^{59}$Co positions. This rule holds for both magnetically soft and hard samples, Fig. 4. This also is reflected in the reversed order of spectral diagrams of MVP action for two type positions.

The reason for this it could be understood also from Fig. 5c, where the reduced echo intensity dependences on the MVP amplitude are shown for $^{55}$Mn and $^{59}$Co positions in the same Co$_2$MnSi sample. It is seen that for these positions amplitude dependences of MVP influence for asymmetric action of MVP differ much stronger for both nuclear types than ones for symmetric MVP influences.

So, the timing diagrams type is defined mainly by the anisotropic part of HF interaction while the degree of suppression of echo signals by MVP strongly depends on the DW mobility.

Let us note also that timing and spectral diagrams in case of half metallic Co$_2$MnSi give a visual picture describing different HF field anisotropies at the $^{59}$Co and $^{55}$Mn sites. Note also that the large difference for DW mobility does not change the type of timing and frequency diagrams.

The shape of frequency diagrams of MVP influence shows the inhomogeneous degree of MVP influence (both symmetric and asymmetric) through a spectrum. The frequency diagram of MVP influence for asymmetric action is arranged below the one for the corresponding symmetric MVP action diagram for anisotropic sites and vice versa for isotropic ones.

The analysis of these diagrams shows that in correspondence with [2] they could play role of additional characteristics of the magnetic materials, as example, for the characterization of nature of the NMR spectra in magnets.

Actually, the appearance of the frequency diagram of MVP action in cobalt is close to the shape of the NMR spectrum. As it is known [10], the NMR spectrum peak at 217 MHz in cobalt corresponds to the nuclei arranged in center of DWs of face-centered cubic (fcc) phase, but at 220 MHz – to nuclei arranged in stacking faults of crystal lattice.

Frequency measurements of MVP influence in cobalt-copper alloy system (Fig. 3) showed a ununiform degree of influence of a MVP on the echo signal in different parts of the

[59]Co NMR spectra in this system. The influence was weakest near 217 MHz corresponding to nuclei located in the DWs centers of fcc phase.

So, the MVP influence technique allows one to make a direct experimental determination of the domain wall center resonances similar to that demonstrated by the NMR spin-echo decay envelope enhanced modulation effect resulting from the application of a small low-frequency alternating (ac) magnetic field [11]. Accordingly model presented in [11], the depth of modulation should be minimal for nuclei located at the DW centers and their edges.

The two techniques are related because the ac magnetic field influence on the echo signal intensity could be mainly approximated by two MVPs because to the first approximation, the amplitude of additional ac magnetic field is important in the process of echo formation only in those instants when RF pulses also affect the sample [12]. But to our opinion the MVPs influence technique is more convenient and direct and offers additional opportunities to characterize magnetic materials.

These considerations holds for NMR peaks arising from a single site in magnetically ordered material containing DWs like Co and its alloys with transition metals.

The appearance of frequency diagrams of MVP influence in Co corresponds to this supposition as the minimal influence is observed for nuclei arranged DWs centers frequencies and maximal – in the range of stacking fault frequencies (~ 220 Hz).

The obtained data for MnSb system suggest that it could be also right for this material and observed NMR spectrum and MVP influence frequency diagrams point to NMR spectrum from nuclei arranged in DW centers and split by the quadrupolar interaction. The characteristic peculiarity of the nuclear spin echo spectrum in the stoichiometric MnSb composition is the presence of well resolved quadrupolar structure in the [55]Mn NMR spectra which is caused by the specific properties of DW structure in these magnets [9]. In particular, into the all range of DW the magnetization vector is perpendicular to the electric field gradient (EFG). As result, the spectral transitions frequencies do not depend on the mutual orientation of the magnetization vector and EFG what stipulates comparatively rare possibility of the observation of the resolved quadrupolar structure in the NMR spectra of polycrystalline magnets. The other specific peculiarity of the investigated system is the relatively small value of HF anisotropy facilitating the interpretation of NMR spectra.

The [55]Mn NMR spectrum of stoichiometrically pure multidomain polycrystalline ferromagnet MnSb is presented by the resonance line in the frequency range of 250-260 MHz splitted by the quadrupolar interaction on five spectral component with widths ~ 1.6 MHz and distances between their centers of the order of 2.0 MHz. At observation of the resonance in the DW of ferromagnet due to the rotation of local magnetization the spread of dipolar fields results in the broadening and splitting of the inhomogeneously broadened NMR lines. The influence of dipolar fields on the resonance spectrum is determined both by the value of dipolar shifts and rotation angles under the action of exciting pulses. Accordingly the assessment [13] the dipole interaction influence in the investigated system is not so large as to disturb significantly the quadrupolar structure of the NMR spectra, but it contributes into the width and shape of the separate components of spectrum splitted by the quadrupolar interaction. At the violated stoichiometry of MnSb it appears the HF shifts near the excess ions [14] resulting in the gradual disappearance of the quadrupolar splitting.

As the consequence of two these factors is apparently the fact that the MVP action is most effective in the intervals between quadrupolar maximums of NMR spectra what could stipulate of the observed shape of the spectral diagram of MVP influence.

### 3. Conclusion

In this work it is carried out the first systematic study of timing and spectral diagrams of MVP influence on TPE in a number of magnets with different anisotropy of HF fields and DW mobility. It is shown that these timing and spectral diagrams are defined by local HF field anisotropy and DW mobility and could be used for additional identification of the nature of NMR lines in multidomain magnetic materials and thereby to improve the resolution capacity of the NMR method for magnets.

### Acknowledgements

The authors thank Professor T.N. Khoperia for kindly providing us with Co TMF samples. The work was supported by the Shota Rustaveli National Science Foundation Short-term Individual Travel Grant 2012_tr_237.